\documentclass[showpacs,amsmath,amssymb]{revtex4}

\newcommand{\aanda}{Astron. Astrophys}

\usepackage{graphicx}
\usepackage{psfrag}

\begin{document}

\title{Revisiting Rotational Perturbations and the Microwave Background}

\author{Viktor G. Czinner}
\email{czinner@rmki.kfki.hu}
\affiliation{Yukawa Institute for Theoretical Physics,\\ 
Kyoto University, Kyoto 606-8502, Japan,\\}
\affiliation{KFKI Research Institute for Particle and Nuclear Physics,\\
Budapest 114, P.O.Box 49, H-1525, Hungary}
\author{M\'aty\'as Vas\'uth}
\email{vasuth@rmki.kfki.hu}
\affiliation{KFKI Research Institute for Particle and Nuclear Physics,\\
Budapest 114, P.O.Box 49, H-1525, Hungary}

\date{\today}

\begin{abstract}
We consider general-relativistic rotational perturbations in homogeneous and isotropic Friedmann -- Robertson -- Walker (FRW) cosmologies. Taking linear perturbations of FRW models, the general solution of the field equations contains tensorial, vectorial and scalar functions. The vectorial terms are in connection with rotations in the given model and due to the Sachs -- Wolfe effect they produce contributions to the temperature fluctuations of the cosmic microwave background radiation (CMBR). In present  paper we obtain the analytic time dependence of  these contributions in a spatially flat, FRW model with pressureless ideal fluid, in the presence and the absence of a cosmological constant $\Lambda$. We find that the solution can be separated into an integrable and a non-integrable part as is the situation in the case of scalar perturbations. Analyzing the solutions and using the results of present observations we estimate the order of magnitude of the angular velocity corresponding to the rotation tensor at the time of decoupling and today. 
\end{abstract}


\maketitle

\section{Introduction}
The importance of the study of possible contributors to the temperature fluctuations of the cosmic microwave background radiation is indisputable. The large angular-scale anisotropy is dominated by the Sachs -- Wolfe effect \cite{SW}, where the first order perturbations in the metric tensor give rise to the appearance of temperature fluctuations in the CMBR. 
 
In present paper we focus on the effects of rotational perturbations in a flat FRW model filled with a single, incoherent, perfect fluid matter source, and we analyze its contributions to the fluctuations both in the absence and the presence of a cosmological constant. This problem certainly has been studied by several authors before, eg. \cite{AM,Barrow,Grish}, however none of these works payed too much attention on the possible analytic solutions in the presence of $\Lambda$. Throughout this paper we follow a systematic treatment in calculating parallely with vanishing and nonvanishing cosmological constant and comparing the obtained results of the calculations. In Section II. we take a general overlook on the $C^{\infty}$ solutions of the perturbed field equations given by Sachs \& Wolfe \cite{SW}, White \cite{White} and Perj\'es et al.~\cite{PVCE}, and the rotation tensor. In the presence of the cosmological constant we use the physical time coordinate $t$ in the calculations -- as the background solution has a closed analytic form only with this time variable -- while in the absence of $\Lambda$ we use the conformal time coordinate $\eta$. In Section III. we calculate the contributions of rotational perturbations to the temperature fluctuations due to the Sachs -- Wolfe effect and analyze the obtained expression. We also give a recipe in this section, how to calculate the anisotropy power spectra of the temperature fluctuations caused by rotational perturbations. In Section IV., using the result of present observations, we give an upper estimation for the order of magnitude of the rotations and the corresponding angular velocity at the time of decoupling and today.

\section{linear perturbations}
In the case of a perfect fluid matter source the Einstein equations are  
\begin{equation} \label{Gab}
G_b^a = (\rho+p ) u^a u_b - p\delta^a_b\ ,
\end{equation} 
with the four velocity $u^a=dx^a/dt$ of the fluid normalized by $u^au_a=1$, $p$ is the pressure, $\rho$ is the energy density and $G_b^a$ is the Einstein tensor. The line element in a spatially flat FRW spacetime with Cartesian coordinate system is
\begin{equation} \label{ivelem}
ds^2=g_{ab}dx^adx^b=a^2(\eta)\eta_{ab}dx^adx^b\ , 
\end{equation} 
where $\eta$ is the conformal time coordinate with the definition $dt=a d\eta$, $a(\eta)$ is the scale factor and $\eta_{ab}$ is the Minkowski metric with the following signature,
\begin{equation}\label{sign}
\eta_{ab}=\eta^{ab}=diag(+1,-1,-1,-1). 
\end{equation}
Here and throughout this paper Roman indices run from 0 to 3, Greek indices from 1 to 3 and we employ units such that $c=8 \pi G=1$. 

The well-known unperturbed solution of the Einstein equations in the case of incoherent matter without cosmological constant is  
\begin{equation}\label{hatter}
p=0\ , \qquad a(\eta) = \frac{2\eta^2}{H_R}\ ,\qquad \rho = \frac{3H_R^2}{\eta^6}\ ,\qquad t=\frac{2\eta^3}{3H_R}\ , \qquad \eta_R=1\ ,
\end{equation} 
where $H_R$ is the present value of the Hubble parameter $H=a^{\prime}/a^2$, the prime denotes the derivative with respect to the conformal time and we chose the $\eta_R=1$ normalization condition. The also well-known background solution of the field equations for incoherent matter in the case of $\Lambda\neq 0$ is   
\begin{eqnarray}  \label{backgr}
p=0\ , \quad a=a_0 {\rm sinh}^{2/3}(Ct+C_0)\ ,\quad
\rho a^3={\cal C}_M\ ,\quad C=\frac{\sqrt{3\Lambda}}{2}\ ,
\quad \eta=\int\frac{dt}{a}\ ,\quad \eta_R=1\ ,\quad C_0=0\ , 
\end{eqnarray}
where $a_0$ is restricted by the normalization condition of $\eta_R$.

Hereafter in this section we follow the perturbation method of Sachs \& Wolfe \cite{SW} and White \cite{White} and we write the perturbed metric $\tilde g_{ab}$ as
\begin{equation} \label{metrika}
ds^2=\tilde g_{ab} dx^a dx^b=a^2(\eta)(\eta_{ab}+h_{ab}) dx^a dx^b\ , 
\end{equation} 
where $h_{ab}$ is the metric perturbation. We use the Minkowski metric $\eta_{ab}$ and its inverse to raise and lower indices of $h_{ab}$ and of other small quantities. In the perturbed spacetime we choose comoving coordinates, where the four velocity of the fluid and the metric perturbation $h_{00}$
are
\begin{equation}  \label{conv}
u^a =\frac{\delta^a_0}{a}\ , \qquad   \quad h_{00}= 0 \ .
\end{equation} 

The general $C^{\infty}$ solution of the perturbed field equations for incoherent matter without cosmological constant is given by Sachs \& Wolfe \cite{SW} and White \cite{White},
\begin{eqnarray} \label{nolsol}
 h_{\alpha \beta}&=& \frac{1}{\eta}\frac{\partial}{\partial \eta} \left(\frac{D_{\alpha \beta}}{\eta} 
  \right) + 2 \left(\frac{\nabla^2}{\eta}-\frac{8}{\eta^3} \right)( C_{\alpha,\beta} 
 +C_{\beta,\alpha} ) +\frac{A_{,\alpha\beta}}{\eta^3}+\eta_{\alpha \beta} B - \frac{\eta^2}{10} 
 B_{, \alpha \beta}\ ,  \\    
 h_{0\alpha } &=& -\frac{2}{\eta^2}\nabla^2 C_{\alpha}\ , \label{nolsol1}\\ 
 \delta\rho&=&\frac{H^2_R}{4}\nabla^2\left(\frac{6A}{\eta^9}-\frac{3B}{5\eta^4}\right)\ ,  \label{drho}  
\end{eqnarray} 
where and throughout this paper comma denotes partial derivative and $A=A(x^\mu)$, $B=B(x^\mu)$ are two scalar functions corresponding to the potentials for density perturbations. The $C_{\alpha}=C_{\alpha}(x^\mu)$ vector function -- satisfying the $C^{\alpha}_{\;\; ,\alpha}=0$ transversality condition -- is related to the perturbed rotation (vorticity) tensor $\omega_{ab}$, see (\ref{rot}). The $D_{\alpha \beta}(x^\mu,\eta)$ transverse, trace-free tensor function represents gravitational waves as the solution of the following wave equation:
\begin{equation}\label{DD} 
\left( \frac{\partial^2}{\partial\eta^2} - \nabla^2 \right) D_{\alpha \beta}= 0\ ,\quad
D_{\alpha \beta} = D_{\beta\alpha }\ , \quad
D_{\alpha \beta}^{\quad, \beta} = 0 \;,\quad D^\mu_{\;\mu}= 0\ ,
\end{equation}
where $\nabla^2\equiv\Delta$ is the standard Laplacian defined by $\nabla^2f=-\eta^{\mu\nu}f_{,\mu\nu}$, for an arbitrary smooth function $f$.

The general $C^{\infty}$ solution of the perturbed field equations for incoherent matter in the case of $\Lambda\neq0$ is given by Perj\'es et al. \cite{PVCE}, 
\begin{eqnarray}\label{hab}
h_{\alpha\beta}&=&S_{\alpha\beta}+\left(\frac{J(t)\nabla^{2}}{a_0C}
-\frac{8Ca_0}{3}\coth(Ct)\right)(C_{\alpha ,\beta }+C_{\beta ,\alpha})+A_{,\alpha\beta}\coth(Ct)
+{\eta_{\alpha\beta}}B-\frac{3I(t)\coth(Ct)}{2^{4/3}a_0^{2}C^{2}}B_{,\alpha\beta}\ , \\   
h_{0\alpha}&=&-\frac{1}{{\rm sinh}^{2/3}(Ct)}\nabla ^{2}C_{\alpha }\ , \label{pertL} \\
\delta \rho &=&\frac{{\cal C}_M}{2a_0^3}\frac{{\rm cosh}(Ct)}{{\rm sinh}^3(Ct)}
\nabla^2\left( A-\frac{3B}{2^{4/3}a_0^2C^2}I(t)\right) \ ,
\end{eqnarray}
where $I(t)$ and $J(t)$ time-dependent amplitudes are the following elliptic integrals,
\begin{equation}\label{J}
I(t)=2^{-2/3}\sqrt{3\Lambda}\int^{t}_0 \frac{{\rm sinh}^{2/3}(C\tau)}{{\rm cosh}^2(C\tau)}
\ {\rm d}\tau\ , \;\;\;\;\;
J(t)=\frac{3}{2^{1/3}}I(t)+3{\rm sinh}^{-1/3}(Ct){\rm cosh}^{-1}(Ct)\ ,
\end{equation}
which can be brought to the Legendre normal form given in the Appendix of \cite{PVCE}. The $S_{\alpha\beta}$ represents gravitational waves analogous to the term containing $D_{\alpha \beta}(x^\mu,\eta)$ in (\ref{nolsol}). We note that in Eqs.(\ref{hab}), (\ref{pertL}) and (\ref{J}) the integration functions $C_{\alpha}$ and $J(t)$ are defined with an opposite sign compared to the original results of Perj\'es et al. in \cite{PVCE}. This sign convention helps us in the following to show the clear correspondence of the forthcoming results derived parallely from the Sachs \& Wolfe and the $\Lambda\neq 0$ solutions.

In the chosen comoving coordinate system, the observers move together with the perturbations of the fluid in the perturbed spacetime. Using the expression for the four velocity $u^a$ given in (\ref{conv}) we can calculate the rotation tensor $\omega_{ab}$ defined by Ehlers \cite{Ehlers}
\begin{equation}
\omega_{ab}=\frac{1}{2}t^{\,c}_{\ a} t^{\,d}_{\ b}(u_{c;d}-u_{d;c})\ ,
\end{equation}
where $t^{\,a}_{\ b}=\delta^a_{\ b}-u^au_b$ and the semicolon denotes the covariant derivative with respect to the perturbed metric $\tilde g_{ab}$. The four velocity $u_a$ is connected to the $h_{0a}$ components of the metric perturbation from which -- with the use of the solution of the field equations in the case of vanishing cosmological constant -- we have the following expression for $\omega_{ab}$  
\begin{equation}\label{rot}
\omega_{\,0 0}=0\ , \qquad \omega_{\,0 \alpha}=0\ , \qquad
\omega_{\alpha\beta}=\frac{1}{H_R}\nabla^2(C_{\beta,\alpha}-C_{\alpha,\beta})\ .
\end{equation}
Repeating the calculations in the case of $\Lambda\neq 0$ we have
\begin{equation}\label{rotl}
\omega_{\,0 0}=0\ , \qquad \omega_{\,0 \alpha}=0\ , \qquad
\omega_{\alpha\beta}=\frac{a_0}{2}\nabla^2(C_{\beta, \alpha}-C_{\alpha,\beta})\ .
\end{equation}
The relations above demonstrate that the perturbed vorticity tensor is completely determined by the $\nabla^2C_{\alpha}$ integration function.

\section{The Sachs -- Wolfe effect from rotational perturbations}
In present section we calculate the contribution of rotational perturbations to the fluctuations of the cosmic microwave background radiation. As a consequence of the Sachs -- Wolfe effect \cite{SW}, the perturbations produce contributions to the temperature fluctuations determined by the following relation 
\begin{equation}
\frac{\delta T}{T}=\frac{1}{2}\int_0^{\eta _R-\eta _E}\left( \frac{\partial
h_{\alpha \beta }}{\partial \eta }e^\alpha e^\beta -2\frac{\partial
h_{0\alpha }}{\partial \eta }e^\alpha \right)  dw \ , \label{deTT1}
\end{equation}
with
\begin{equation}
\eta=\eta_R-w\ , \qquad x^{\beta}=e^{\beta}w\ , \qquad e^{\beta}e_{\beta}=-1\ , \qquad e^{\beta}=const.
\end{equation}
Here the vector $e^{\alpha}$ represents the spatial direction of the light signal as seen by an observer moving together with the fluid and the $E$ subscript denotes the instant of time of the emission. Applying the following identity for an arbitrary smooth function $f=f(x^a)$
\begin{equation}
f_{,a}\frac{dx^a}{dw}dw=f_{,\alpha }e^\alpha dw-f^{\prime }dw \ , 
\end{equation}
and calculating only with rotational perturbations (terms containing $C_{\alpha}$ in the perturbed solutions), (\ref{deTT1}) can be written in the following general form, 
\begin{eqnarray}\label{gf}
\frac{\delta T}{T}=\left[\left(\phi(\eta)-\psi(\eta)\nabla^2
\right)C_{\alpha}e^{\alpha}\right]_0^{\eta_R-\eta_E}-\int_0^{\eta_R-\eta_E}\xi(\eta) C_{\alpha}e^{\alpha} dw\ ,
\end{eqnarray}
were the $\phi$, $\psi$ and $\xi$ time dependent amplitudes are the following in the cases of vanishing and nonvanishing cosmological constant respectively:
\begin{eqnarray}
\phi(\eta)&=&\frac{48}{\eta^4}\  ,\qquad \qquad \qquad \quad\, \psi(\eta)=\frac{2}{\eta^2}\  ,\qquad \qquad \qquad \ \, \xi(\eta)=\frac{192}{\eta^5}\ , \\
\phi_{\Lambda}(t)&=&\frac{8C^2a_0^2}{3\sinh^{4/3}(Ct)}\  ,\qquad\ \! \psi_{\Lambda}(t)=\frac{1}{\sinh^{2/3}(Ct)}\  ,
\qquad\ \! \xi_{\Lambda}(t)=\frac{32C^3a_0^3}{9}\frac{\cosh(Ct)}{\sinh^{5/3}(Ct)}\ .
\end{eqnarray}

In FIG.1 we have plotted the $\phi$, $\psi$ and $\xi$ time dependent amplitudes on a logarithmic scale. The shape of the curves show that the presence of the cosmological constant results only about a $\sim 1/3$ factor difference from the single fluid model at each functions. This means that the fluctuations represented by rotational perturbations in the microwave background are $\sim 3$ times smaller in the presence of $\Lambda$.
It is also remarkable that the magnitude of the amplitudes are decreasing exponentially since the time of decoupling, and the present day contributions are negligible comparing to early epochs. Except the integrated term in (\ref{gf}) -- whose contribution is growing in time -- the contributions of the first two terms are the difference of their values at the time of decoupling and today.   Therefore the dominant parts come from the time of decoupling and -- as the present values are many orders of magnitude smaller -- these terms produce the only relevant contributions to the temperature fluctuations that could be measured today.       
\begin{figure}
\psfrag{t}{$\!\!\!\!\!\!t_{dec}$}
\psfrag{T}{\!\!$t_{dec}(\Lambda)$}
\psfrag{r}{$\!\!\!\!t_{R}$}
\psfrag{R}{$\!\!t_R(\Lambda)$}
\psfrag{a(x)}{\!\!\!\!\!{\footnotesize$\psi(t)$}}
\psfrag{b(x)}{\!\!\!\!\!{\footnotesize$\psi_{\Lambda}(t)$}}
\psfrag{c(x)}{\!\!\!\!\!{\footnotesize$\phi_{\Lambda}(t)$}}
\psfrag{d(x)}{\!\!\!\!\!{\footnotesize$\phi(t)$}}
\psfrag{e(x)}{\!\!\!\!\!{\footnotesize$\xi(t)$}}
\psfrag{f(x)}{\!\!\!\!\!\!{\footnotesize$\xi_{\Lambda}(t)$}}
\psfrag{t (year)}{$t\,(year)$}
\psfrag{1e+06}{$10^6$}
\includegraphics{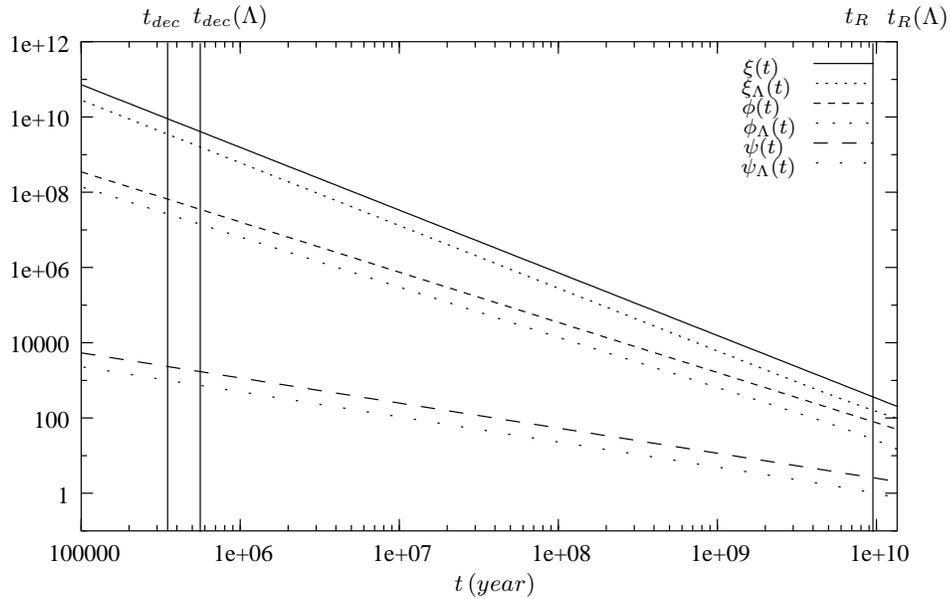}
\caption{\label{fig1} The time dependence of the $\phi(t)$, $\psi(t)$ and $\xi(t)$ functions in the presence and the absence of $\Lambda$. The subscript $\Lambda$ denotes the corresponding functions in the model with cosmological constant. The vertical lines belong to the following events: $t_{dec}(\Lambda)$ and $t_{dec}$ denote the time of decoupling in the models with and without cosmological constant, while the $t_{R}(\Lambda)$ 
and $t_{R}$ denote the present age of the universe in the two models respectively.}
\end{figure}

In the remainder of this section we present the recipe of the calculations how to obtain the anisotropy spectra of the fluctuations generated by rotational perturbations. As a first step, following Bardeen's covariant formalism, we replace the $C_{\alpha}(x^{\beta})$ perturbation function with a single mode of its decomposition into the vector harmonics $Q_{\alpha}^{(m)}$ of the Helmholtz equation \cite{Bardeen}
\begin{eqnarray}
C_{\alpha}(x^{\beta}) = \!\!\!\sum_{m=-1}^1 C^{(m)}_k \,Q_{\alpha}^{(m)}\ ,
\end{eqnarray}
where the subscript $k$ denotes the wavenumber dependence of the coefficients $C^{(m)}_k$ and
\begin{eqnarray}
{ Q^{(0)} } &=& { \exp( i {\bf k} \cdot {\bf x}) } \,,  \nonumber\\
Q_{\alpha}^{(0)} & = & -k^{-1} Q^{(0)}_{\,,\alpha}\,,\\
{ Q_{\alpha}^{(\pm 1)} }
&=& \frac{-i}{\sqrt{2}} ({\bf e}_1 \pm i {\bf e}_2)_{\alpha} 
{ \exp(i {\bf k}\cdot {\bf x})\,.\nonumber }
\end{eqnarray}
The spatial unit vectors ${\bf e}_1$ and ${\bf e}_2$ are spanning the plane perpendicular to ${\bf k}$.

Applying this expansion, the integrated term in (\ref{gf}) can be written in the following form in the case of vanishing cosmological constant 
\begin{eqnarray}
ISW&=&-\frac{192i}{\sqrt{2}}\sum_{\pm1}C_k^{(\pm1)}({\bf e}_1\pm i{\bf e}_2)_{\alpha}e^{\alpha}
\int_0^{1-\eta_E}\frac{e^{ik_{\beta}e^{\beta}w}}{(1-w)^5}dw\ .
\end{eqnarray}
Using the $k_{\beta}e^{\beta}\equiv k\cos(\zeta)=n$ replacement one can evaluate this ISW integral in a closed analytic form as follows,
\begin{eqnarray}
ISW=&-&\frac{8i}{\sqrt{2}\eta^4_E}\sum_{\pm 1}C^{(\pm1)}_k({\bf e}_1\pm i{\bf e}_2)_{\alpha}e^{\alpha}
\left[e^{in(1-\eta_E)}\left(6-2in\eta_E-n^2\eta_E^2+in^3\eta_E^3\right)\right. \nonumber\\
&+&\left.\eta_E^4n^4e^{in}\left(\Gamma(0,in\eta_E)-\Gamma(0,in)\right)-\eta_E^4
\left(6-2in-n^2+in^3\right)\right]\ .
\end{eqnarray}
Repeating the calculations in the case of $\Lambda\neq0$, the integrated term in (\ref{gf}) takes the following shape
\begin{eqnarray}\label{iswl}
ISW(\Lambda)&=&\frac{16iC^2a_0^2}{3\sqrt{2}}\sum_{\pm1}C^{(\pm1)}_k({\bf e}_1\pm i{\bf e}_2)_{\alpha}e^{\alpha}\int_{y_R}^{y_{E}}\frac{e^{in\left(1+\chi(y)\right)}}{y^3}dy\ ,
\end{eqnarray}
where $y=a/a_0=\sinh^{(2/3)}(Ct)$ and
\begin{equation}
\chi(y)=-\frac{3}{2Ca_0}\int_0^y\frac{dy^{\prime}}{\sqrt{y^{\prime}+y^{\prime 4}}}=\sqrt{\frac{3}{2}}\frac{\sqrt{3-\sqrt{3}i}}{Ca_0}
F\left(\sqrt{\frac{(3+\sqrt{3}i)y}{2(y+1)}}\left\vert\frac{i-\sqrt{3}}{2}\right. \right) \ . 
\end{equation}
The ISW($\Lambda$) integral in (\ref{iswl}) can not be evaluated analytically thus numerical integration needs to be used.

To be able to obtain the anisotropy power spectra generated by rotational perturbations, one has to evaluate the following well-known integral
\begin{equation}\label{cL}
{2 \ell + 1 \over 4\pi} C_\ell = {V \over 2\pi^2} \int_0^\infty {d k \over k}
k^3 {|\Theta_\ell(\eta,k)|^2 \over 2\ell+1}\quad  ,
\end{equation}
where $C_{\ell}$ is the anisotropy momenta and $\Theta_{\ell}(k,\eta)$ is identical with $\delta T/ T$ in (\ref{gf}) -- in the decomposition by vector harmonics of the Helmholtz equation presented above. 

As the last step, one has to take the square of the absolute value of $\delta T/T$ given in (\ref{gf}). To be able to evaluate the integral expression of $C_{\ell}$, one needs to have information about the spectrum of $C_k$ which might be originated from predictions of models describing the universe before the decoupling era. As an example this has been modelled by Grishchuk in \cite{Grish}, where he derives the angular variation of the microwave background radiation from rotational perturbations of quantum-mechanical origin. 

\section{Estimations of the magnitude of rotations}\label{omega}

As an application we estimate the magnitude of the angular velocity generated by rotational perturbations using the results of present observations of the CMBR fluctuations. It is well-known from the analysis of recent measurements \cite{dipole} that the dipole component of the temperature fluctuations has the largest amplitude with the order of magnitude $\sim 10^{-3}$. This term certainly contains many contributions, however one of these originates from the effect of rotational perturbations. As we have shown in (\ref{gf}) this contribution can be generally separated into the following three components: one proportional to $C_{\alpha}$, one proportional to $\nabla^2C_{\alpha}$ and an integrated term proportional also to $C_{\alpha}$. Since the rotation tensor  is completely determined by the $\nabla^2C_{\alpha}$ function, we must estimate the order of magnitude of the $\nabla^2C_{\alpha}$ perturbation to be able to obtain the magnitude of the corresponding angular velocity vector {\mbox{\boldmath$\omega$}}$(t,x^{\alpha})$. Applying the results of present observations of the dipole component, we can make an estimation on the upper bound of the rotations with the following procedure.

The corresponding angular velocity  deduced from the vorticity tensor is
\begin{equation}
\omega^{\alpha}=\frac{1}{2\sqrt{\gamma}}\varepsilon^{\alpha\beta\delta}\omega_{\beta\delta}\ ,
\end{equation}
where $\gamma=-{\rm det}(g_{\alpha\beta})$ and $\varepsilon^{\alpha\beta\delta}$ is the 3-dimensional totally antisymmetric Levi-Civita symbol. Having done the calculation for the square root of $\omega^2$, we obtain the expression 
\begin{eqnarray}\label{kappa}
\omega(t,x^{\alpha})=w(t)\kappa(x^{\alpha}) \quad \mbox{with} \quad \kappa(x^{\alpha})=\left[(\nabla^2C_{\alpha , \beta})^2-\nabla^2C^{\alpha , \beta}\nabla^2C_{\beta , \alpha}\right]^{1/2},
\end{eqnarray}
and the $w(t)$ time dependent amplitude is the following in the cases of vanishing and nonvanishing cosmological constant respectively,
\begin{eqnarray}
w(\eta)&=&\frac{H_R}{4\eta^4}, \qquad  w_{\Lambda}(t)=\frac{1}{2a_0\sinh^{(4/3)}(Ct)}\ .
\end{eqnarray}

As an upper limit, in accordance with the measurements, we can say that none of the terms in (\ref{gf}) can be larger than the order of magnitude $10^{-3}$, so this must also be valid for the $\nabla^2C_{\alpha}$ term. In the previous section we have shown that the relevant contribution of this term comes from its value at the time of decoupling, where the magnitude of its time dependent amplitude $\phi(t_{dec})$ is known. This makes the restriction to $\nabla^2C_{\alpha}$ that its order of magnitude can not be larger than $10^{-6}$. If we assume that the first derivative $\nabla^2C_{\alpha , \beta}$ has the same order of magnitude, we can make the estimation for $\kappa(x^{\alpha})$ in (\ref{kappa}) that is $\kappa(x^{\alpha})\sim 10^{-6}$. Applying this assumption we obtain the following order of magnitudes for the values of the angular velocity at the time of decoupling and today in the presence and the absence of $\Lambda$:

\begin{equation}
\omega(t_{dec},x^{\alpha}),\  \omega_{\Lambda}(t_{dec},x^{\alpha})\sim 10^{-11}\,/year\ ;\qquad 
\omega(t_{R},x^{\alpha}),\  \omega_{\Lambda}(t_{R},x^{\alpha})\sim 10^{-17}\,/year\ .
\end{equation}

It shows that the presence of the cosmological constant doesn't change the order of magnitude of the angular velocity, however in a more accurate calculation there might be a multiplying factor $(<10)$ between the two models similar to the results obtained for $\delta T/T$ in the previous section.

The present value of $\omega$ can be associated with the rotation of the universe, and the obtained order of magnitudes are in good agreement with the results of \cite{Barrow}, where the estimated value for the ratio  $\omega(t_R)/H_R<10^{-5}$ in a flat model. 

\section{Conclusions}
We have considered the effects of rotational perturbations of FRW models on the cosmic microwave background radiation and its connection to the amplitude of the rotation of the universe. We presented the general form of the relative temperature fluctuations induced by rotations and obtained analytic expressions for all the time dependent coefficients arising both in the presence and the absence of a cosmological constant. It is found that the $\phi$ and $\psi$ transfer functions of the rotational Sachs--Wolfe effect, and the $\xi$ weighing function of the rotational integrated Sachs--Wolfe effect are decaying exponentially, and there is a $\sim 1/3$ factor between their values in the models with zero and nonzero cosmological constant. We presented the general method of calculating the anisotropy power spectra of the rotational ISW term, and using the results of recent observations for the dipole component of the microwave background, we made an upper estimation for the magnitude of the rotation of the universe at the time of decoupling and today. It is a remarkable result that the value of the corresponding angular velocity at the time of decoupling is close to the order of magnitude of the angular velocity of a typical galaxy cluster today \cite{cluster1,cluster2}. This interesting coincidence might deserve attention in future investigations.

\begin{acknowledgments}
We thank \'Arp\'ad Luk\'acs and L\'aszl\'o B. Szabados for useful conversations and for their help in preparing the manuscript. We are also thankful to Lajos Bal\'azs for his valuable help to find our way in the  astronomical literature. This work was supported by JSPS P06816 and OTKA No. F049429 grants. 
\end{acknowledgments}
\appendix*
\section{physical parameters}
We present the list of physical parameters that have been used throughout the calculations obtaining numerical values.  
\begin{eqnarray}
H_R&=&70\,km\,s^{-1}Mpc^{-1}\ ;\\
\Omega&=&\Omega_{Matter}+\Omega_{\Lambda}=1\ ;\quad \Omega_{Matter}=0.3\ ,\quad\Omega_{\Lambda}=0.7\ ;\\
\Lambda&=&3H_R^2\Omega_{\Lambda}/c^2=1.2\times10^{-52}\,m^{-2}\ ;\\
C&=&\sqrt{3\Lambda}/2=2.85\times10^{-18}\,s^{-1}\ ;\\
a_0&=&1.1\times10^{18}\,s\ ;\\
z_{dec}&\simeq& 1100\ ;\\
\eta_R&=&1\ .
\end{eqnarray}

\end{document}